\documentclass[iop,apjl,tighten]{emulateapj}
\usepackage{graphicx,enumitem}

\usepackage{color}

\newcommand{\Msun}{{M_\odot}}
\newcommand{\Ecrit}{{E_{\rm crit}}}
\usepackage{soul}   

\shorttitle{The Sausage Globular Clusters}
\shortauthors{Myeong et al.}

\begin{document}

\title{The \lq Sausage' Globular Clusters}

\author{G.~C. Myeong$^{1}$, N.~W. Evans$^1$, V. Belokurov$^1$,
  J.~L. Sanders$^1$ and S.~E. Koposov$^{1,2}$}

\affiliation{$^1$Institute of Astronomy, University of Cambridge,
  Madingley Road, Cambridge CB3~0HA, United Kingdom\\ $^2$McWilliams
  Center for Cosmology, Department of Physics, Carnegie Mellon
  University, 5000 Forbes Avenue, Pittsburgh PA 15213, USA}

\begin{abstract}
  The {\it Gaia Sausage} is an elongated structure in velocity space
  discovered by Belokurov et al. (2018) using the kinematics of
  metal-rich halo stars.  It was created by a massive dwarf galaxy
  ($\sim 5 \times 10^{10} \Msun$) on a strongly radial orbit that
  merged with the Milky Way at a redshift $z\lesssim 3$. We search for
  the associated {\it Sausage Globular Clusters} by analysing the
  structure of 91 Milky Way globular clusters (GCs) in action space
  using the {\it Gaia} Data Release 2 catalogue, complemented with
  {\it Hubble Space Telescope} proper motions.  There is a
  characteristic energy $\Ecrit$ which separates the {\it in situ}
  objects, such as the bulge/disc clusters, from the accreted objects,
  such as the young halo clusters. There are 15 old halo GCs that have
  $E > \Ecrit$.  Eight of the high energy, old halo GCs are strongly
  clumped in azimuthal and vertical action, yet strung out like beads
  on a chain at extreme radial action.  They are very radially
  anisotropic ($\beta \sim 0.95$) and move on orbits that are all
  highly eccentric ($e \gtrsim 0.80$). They also form a track in the
  age-metallicity plane distinct from the bulk of the Milky Way GCs
  and compatible with a dwarf spheroidal origin.  These properties are
  consistent with GCs associated with the merger event that gave rise
  to the Gaia Sausage.
\end{abstract}


\section{Introduction}

There are multiple and striking pieces of evidence for the existence
of a massive ancient merger which provides the bulk of the stars in
the inner halo of the Milky Way galaxy.  For example, the radial
density profile of the stellar halo shows a dramatic break at around
30 kpc in tracers such as RR Lyrae and blue horizontal branch
stars~\citep[e.g.,][]{Wa09,De11}. \citet{De13} argued that this could
be interpreted as the last apocentre of a massive progenitor galaxy
accreted between 8 and 10 Gyr ago. \citet{My18a} showed that the
kinematics of metal-rich halo stars ($-1.9 <$ [Fe/H] $< -1.1$) betray
extensive evidence of recent accretion using the SDSS-{\it Gaia}
catalogue. The variation in Oosterhoff classes of RR Lyraes with
radius \citep{Be18a} similarly shows evidence that the bulk of the field
RRab is provided by a single massive progenitor. Finally, \citet{Be18}
demonstrated that the shape of the velocity ellipsoid of the inner
metal-rich stellar halo is highly non-Gaussian and sausage-shaped.
They interpreted this {\it Gaia Sausage} as evidence that two thirds of
the local stellar halo could have been deposited via the disruption of
a massive ($\gtrsim 10^{10} \Msun$) galaxy on a strongly radial orbit
between redshift $z=3$ and $z=1$. Although identified in the SDSS-{\it
  Gaia} catalogue, recent investigations by \citet{Ha18} with the new
{\it Gaia} Data Release 2 (DR2) catalogue~\citep{GaDR2} support the
original hypothesis.  If so, then this beast must have brought with it
a population of globular clusters (GCs), now dispersed in the inner
halo. After all, the similarly massive Sagittarius galaxy (Sgr) is now
known to have brought at least 4 and possibly 7 GCs with
it~\citep[e.g.,][]{Fo10,So18}.

The main aim of this {\it Letter} is to search for the {\it Sausage
Globular Clusters}. The identification of objects accreted in the same
merger event is easiest in action space. Actions have the property of
adiabatic invariance, so that they stay approximately constant when
changes in the potential occur slowly~\citep[e.g.,][]{Go80,Bi82}.
Globular clusters accreted in the same event are identifiable as
clumped and compact substructures in action space (as is indeed the
case for the 4 Sgr GCs -- Terzan 7, Terzan 8, Arp 2, Pal 12).
Historically, actions were cumbersome to calculate, but recent
theoretical advances have transformed the
situation~\citep[e.g.,][]{Bi12,Sa16}. The power of actions has
recently been demonstrated by the identification of the tidal
disgorgements of $\omega$~Centauri~\citep{My18b}.  Here, we display
the Milky Way globular clusters in action space using a realistic
Galactic potential comprising flattened stellar and gas discs, halo
and bulge~\citep{Mc17} with the specific aim of identifying the
{\it Sausage Globular Clusters}.

\begin{figure*}
  \includegraphics[width=180mm]{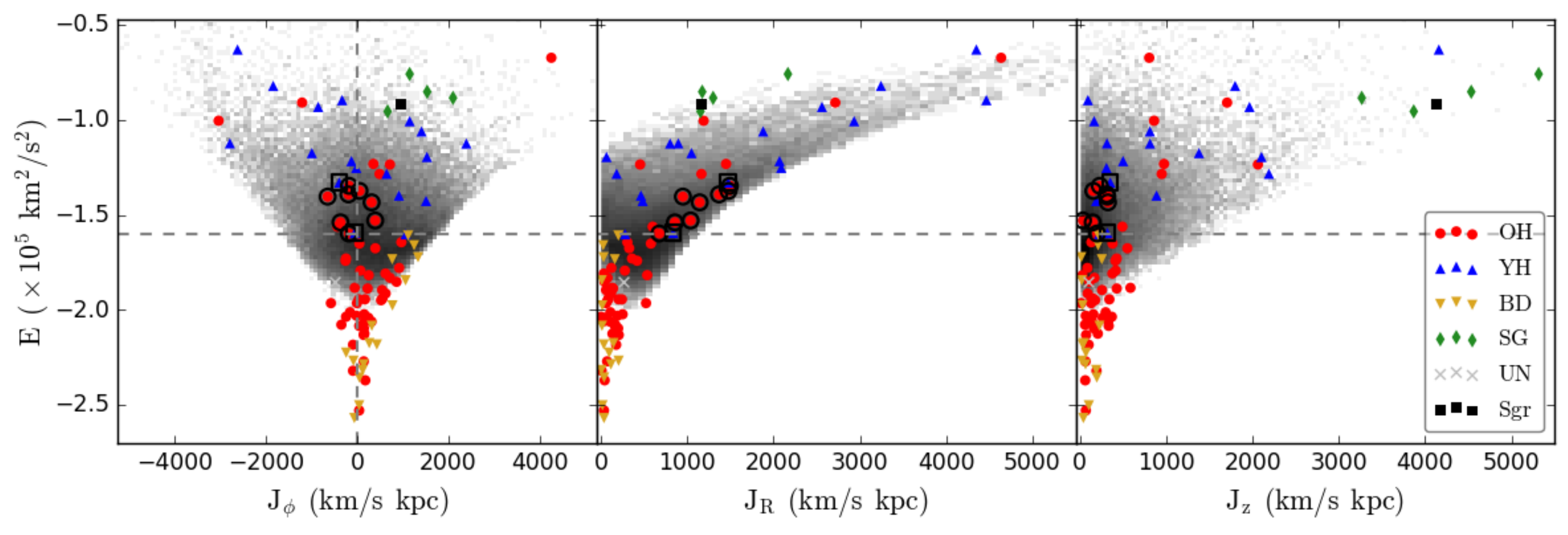}
  \caption{The distribution of globular clusters (GCs) in
    energy-action space or ($J_\phi,E$), ($J_R, E$) and ($J_z,E$)
    space. The grey-scale background shows the halo main-sequence
    turn-off (MSTO) stars from \citet{My18a} as a comparison.  There
    are 75 GCs with {\it Gaia} DR2 proper motions and a further 16
    with {\it Hubble Space Telescope} proper motions; 53 old halo
    (OHs, red circles), 17 young halo (YHs, blue triangles), 16
    bulge/disc (BDs, yellow triangles), and 4 Sgr GCs (SG, green
    diamonds) together with 1 of unknown classification (grey cross).
    The Sagittarius galaxy (Sgr) is also marked as a black filled
    square.  The vertical dashed line marks the division between
    prograde ($J_\phi>0$) and retrograde ($J_\phi <0$). The horizontal
    dashed line signifies the characteristic energy above which all
    the YHs lie, and below which all the BDs lie. The eight OH
    globular clusters whose symbols are enclosed by black open circles
    are grouped together in ($J_\phi, E$) and ($J_z, E$), whilst in
    ($J_R,E$) they are stretched out close to the boundary of $J_R$ at
    corresponding energy (as judged from the MSTOs). They are the
    Sausage GCs. The 2 YHs enclosed with black open squares form
    an extended selection that may also be related. They have horizontal
    branch morphology similar to OHs, and have similar actions.
    \label{fig:actionspace}}
\end{figure*}
\begin{figure*}
  \includegraphics[width=180mm]{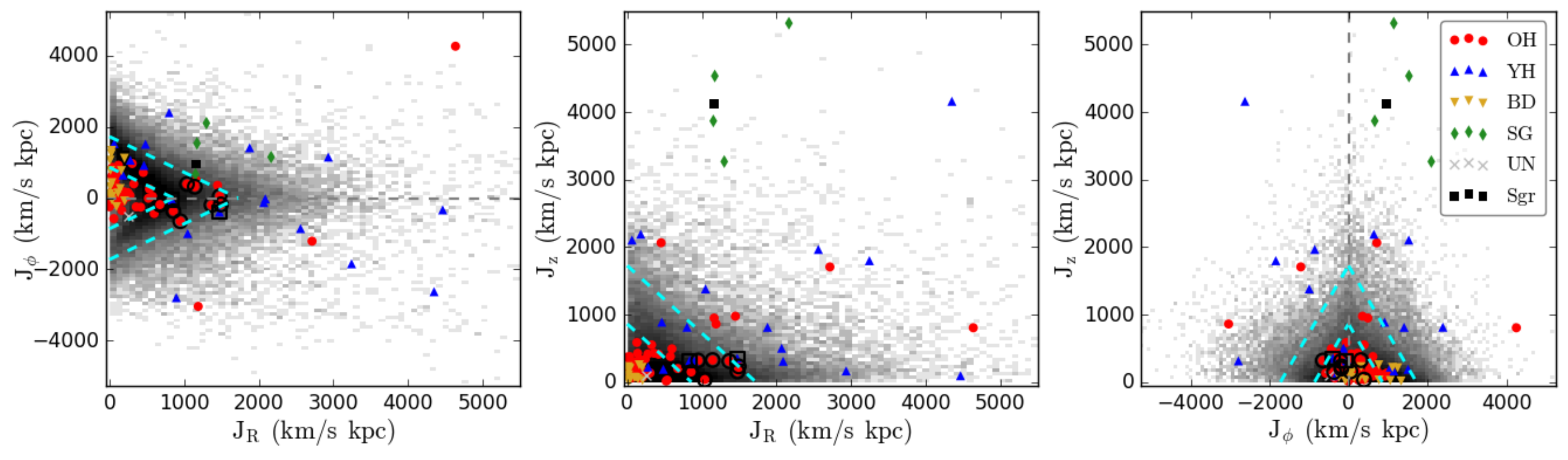}
  \caption{The same data as Fig.~\ref{fig:actionspace}, but now in
    action space. The Sausage GCs form an extended sequence in $J_R$,
    but are tightly clustered in $J_\phi$ and especially $J_z$. Again
    black circles enclose probable members, black open squares
    possibles; red circles are OHs, blue triangles are YHs, yellow
    triangles are BDs, green diamonds are SGs and grey cross is
    unknown.  The black filled square is Sgr itself.  The grey dashed
    line marks $J_\phi=0$.  The two cyan dashed lines mark two
    constant-energy surfaces projected onto the principal planes to
    provide a rough idea of the action space morphology \citep[see
      e.g., Figure 3.25 of][]{BT08}.
\label{fig:acttwo}}
\end{figure*}
\begin{figure}
  \includegraphics[width=75mm]{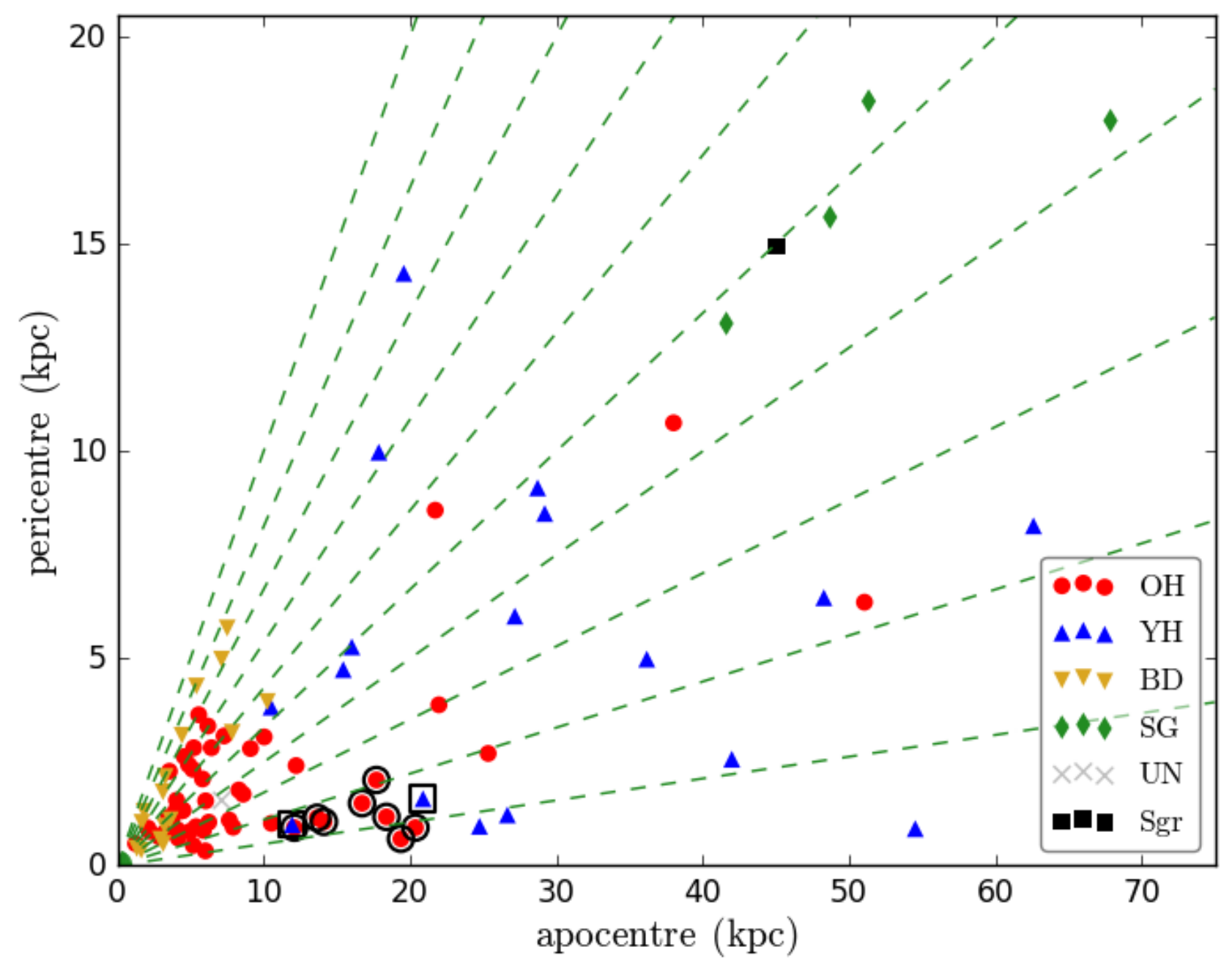}
    \includegraphics[width=75mm]{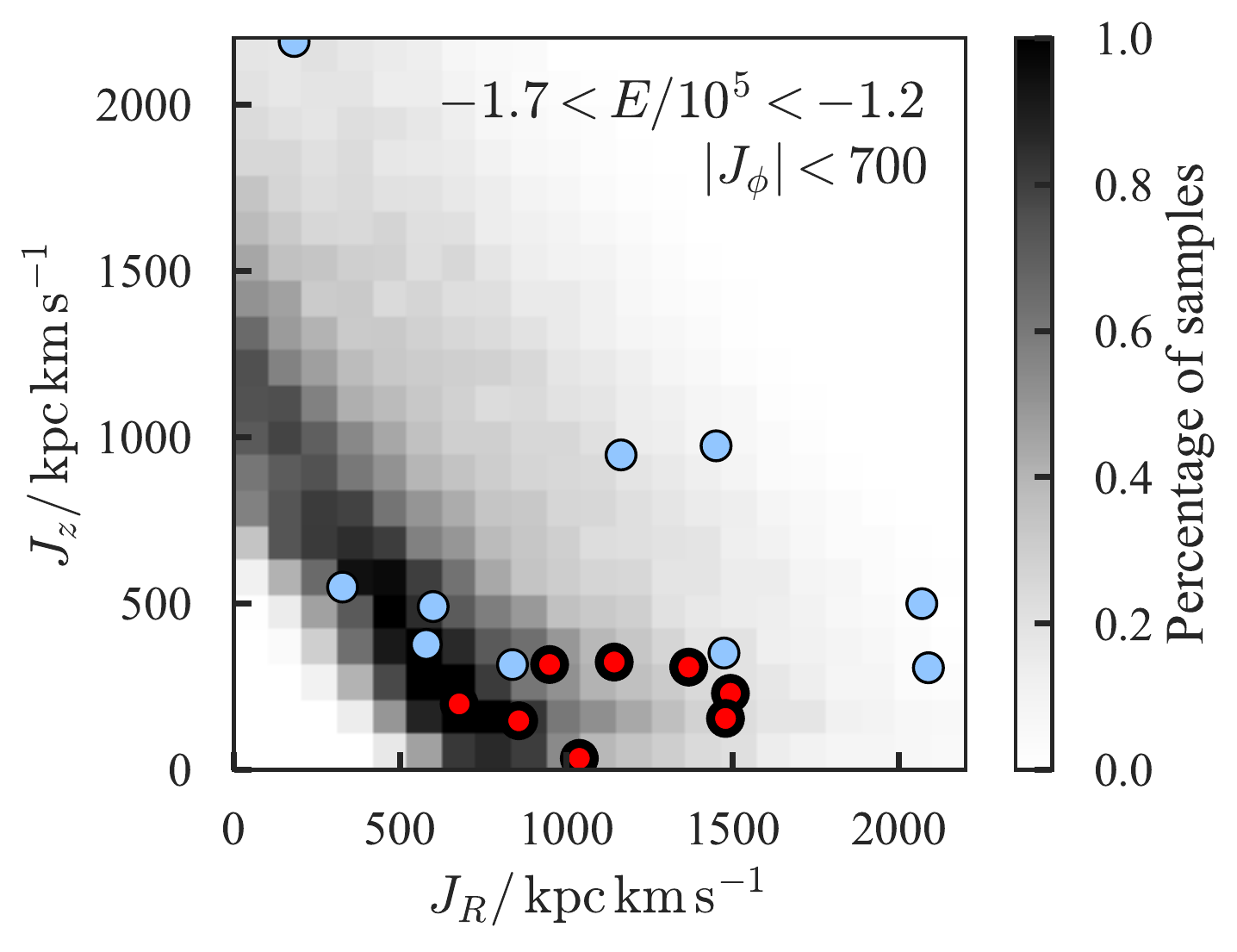}
  \caption{Upper Panel: Apocentres and pericentres of the GCs, colour
    coded according to old halo (red circles), young halo (blue
    triangles), bulge/disc (yellow triangles), Sgr GCs (green
    diamonds), and unknown (grey cross). Sagittarius galaxy (Sgr) is
    also marked (black filled square). Lines of constant eccentricity
    from 0 to 0.9 in steps of 0.1 are shown in green. Note the Sausage
    GCs (black open circles as probables and open squares as
    possibles) all have eccentricity $\gtrsim 0.80$. Lower Panel: {\it
      Gaia} Selection Effects.  The grey pixels show the distribution
    of samples in action space of GCs at the observed locations of
    GCs, but with velocities randomly drawn from isotropic Gaussians
    with velocity dispersion $\sigma = 130$ kms$^{-1}$.  Only samples with $-1.7<E/10^5 \,($km$^2$/s$^2)<-1.2$ and
    $|J_{\phi}|<700$ kms$^{-1}$kpc are shown. Although
    there is a weak bias to low $J_z$, it is clear that {\it Gaia}
    could have detected objects at high $J_z$ in this energy range if
    they existed. The actual locations of the Sausage GCs (red) and
    other GCs (pale blue) with $-1.7<E/10^5 \,($km$^2$/s$^2)<-1.2$ and
    $|J_{\phi}|<700$ kms$^{-1}$kpc are superposed.
\label{fig:ecc}}
\end{figure}
\begin{figure*}
  \includegraphics[width=180mm]{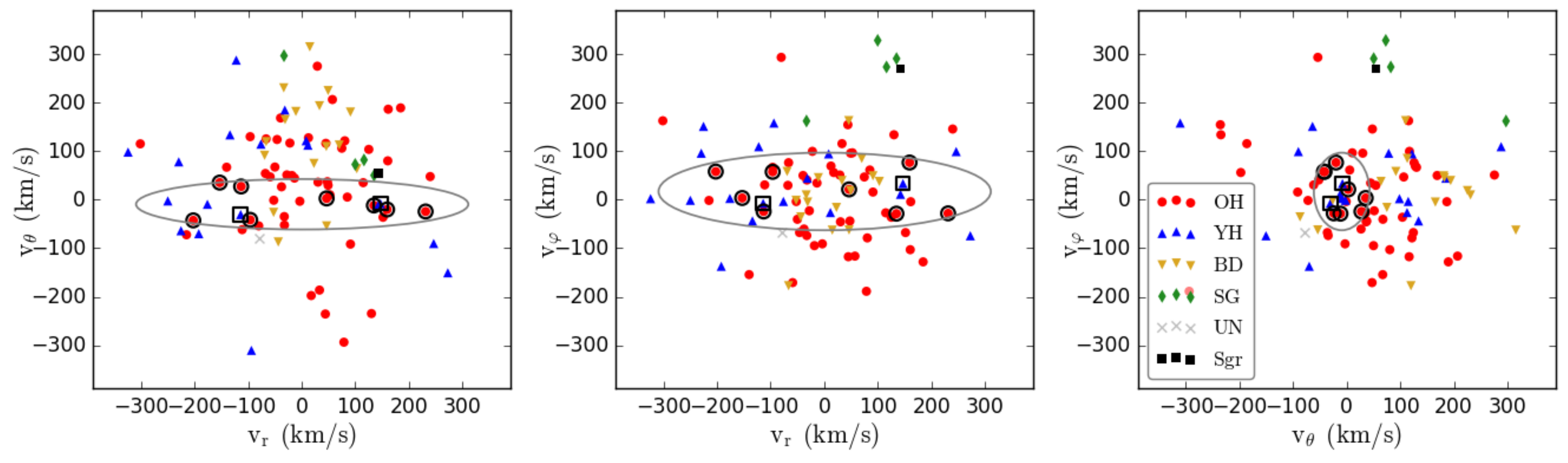}
  \caption{The velocity distribution of the GC sample, resolved with
    respect to spherical polar coordinates ($v_r, v_\theta, v_\phi$).
    The Sausage GCs are marked with their customary black open circles
    (probables) and open squares (possibles).  Their extreme radial
    anisotropy is illustrated by the superposed ellipses with semiaxes
    given by the velocity dispersion in each coordinate. This plot
    should be compared with Fig. 2 of \citet{Be18}, which shows the
    sausage-like velocity distributions of main-sequence turn-off
    stars in the SDSS-{\it Gaia} catalogue.
\label{fig:velocities}}
\end{figure*}

\section{The Globular Clusters in Action Space}

The Milky Way globular clusters are a disparate group: some were
formed {\it in situ} in the Milky Way, some acquired by the engulfment
of dwarf galaxies. A classification was introduced by \citet{Zi93}, in
which globular clusters are divided into bulge/disc, old halo and
young halo on the basis of cluster metallicity and horizontal branch
morphology. The bulge/disc systems are concentrated in the Galactic
bulge and inner disc, whilst the old halo clusters are predominantly
in the inner halo. They are mostly believed to have been formed in the
Milky Way, though $\sim$\,15--17 per cent might have been accreted.
The young halo clusters can extend to large radii and are all believed to
have been accreted~\citep[see e.g.,][]{MG04, MB05}.

The combination of observables from the \citet{GaDR2}, \citet{So18}
and \citet[2010 edition]{Ha96} allows us to obtain full
six-dimensional information for 91 globular clusters (out of a total
of $\sim 150$ in the Galaxy).  To convert from observables to the
Galactic rest-frame, we use the circular speed of 232.8 kms$^{-1}$ at
the Sun's position of 8.2 kpc, consistent with the \citet{Mc17}
potential, whilst for the Solar peculiar motion we use the most recent
value from \citet{Sc10}, namely $(U,V,W) = (11.1, 12.24, 7.25)$
kms$^{-1}$. These values differ from those used by the \citet{GaDR2}
or \citet{Po18}, so there are small differences in quantities such as
apocentres and eccentricities.  We use the numerical method of
\citet{Bi12} and \citet{Sa16} to compute the action variables of each
globular cluster ($J_R, J_\phi, J_z$). Globular clusters associated
with the Gaia Sausage must lie on highly radial orbits, and so have
low $J_\phi$ and $J_z$, but very large $J_R$.  The uncertainty in
proper motions is the main contributor to the median (total) velocity
error of $\sim 9$ kms$^{-1}$. This leads to median errors in the
actions of $\sim 10 \%$, and so features in action space are robust
against uncertainties. Fig.~\ref{fig:actionspace} presents the
distribution of globular clusters in energy and action space, while
Fig.~\ref{fig:acttwo} presents the projections onto the principal
planes of action space. In both cases, we also show as grey pixels the
distribution of main sequence turn-off stars (MSTOs) from
\citet{My18a}. This is to give an idea of the range in action at any
energy level occupied by the stellar halo. Both plots are colour-coded
according to the conventional classification from \citet{MB05}: red
circles mark the old halo globular clusters (OHs), the blue triangles
the young halo globular clusters (YHs), yellow triangles the
bulge/disc ones (BDs) and green diamonds the Sagittarius GCs (SG). The
Sagittarius dwarf (Sgr) is also marked as a black filled square. The
young halo globular clusters all lie above a critical energy of
$\Ecrit = -1.6 \times 10^5$ km$^2$s$^{-2}$.  The bulge/disc globulars
all lie below this critical energy.  We regard the identification of
this critical energy $\Ecrit$ as a reference level. Though the value
of $\Ecrit$ does depend on potential, the existence of a critical
energy level is robust -- it is the value of the most bound YH
cluster.  We argue that globular clusters with comparable or higher
energy are all accreted from dwarf galaxies.

The bulge/disc and old halo clusters form tracks in
Figs.~\ref{fig:actionspace} and ~\ref{fig:acttwo}.  We can see that
the bulge/disc clusters branch out towards positive $J_\phi$, while
maintaining low $J_R$ and low $J_z$ values, as befit disc
orbits. They are entirely limited to $E \le \Ecrit$.  For the old halo
clusters, we can see a similar branching towards positive $J_\phi$ at
low energy ($E < \Ecrit$). The low energy old halo clusters are all
concentrated at low $J_R$. In fact, there is a branch with $J_R$
decreasing with increasing energy for the low energy old halo clusters
-- as is also the case for the MSTOs in \citet{My18a}. In the
($J_z,E$) plane, the old halo clusters seemingly break up into two
separate branches at low energy, though it is unclear whether this is
caused by dynamical or selection effects.

There are 15 old halo clusters above the critical energy ($E \gtrsim
\Ecrit$). Their azimuthal action $J_\phi$ distribution is narrower
than the low energy ones. It resembles the tips of the \lq
diamond-like' contours seen in the distribution of MSTOs in the
metal-rich halo~\citep{My18a}. Also, the radial action $J_R$
distribution of high energy old halo clusters is extremely
distended. Most of them have high radial action, tracing out a
structure similar to the picture of the metal-rich halo.

\begin{table}
  \caption{The kinematic properties of the 8 probable and 2 possible
    Sausage GCs. The Galactic rest frame velocity in spherical polars,
    the actions in cylindrical polars, the energy and orbital
    eccentricity $e = (r_{\rm apo} - r_{\rm peri})/(r_{\rm apo} + r_{\rm
      peri})$ are all given.}
 \begin{center}
   \begin{tabular}{lrrrc}
 \hline \hline
 \multicolumn{1}{c}{Name} &
 \multicolumn{1}{c}{$(v_r,v_\theta,v_\varphi)$} &
 \multicolumn{1}{c}{$e$} &
 \multicolumn{1}{c}{$(J_R,J_\phi,J_z)$} &
 \multicolumn{1}{c}{E} \\

 \multicolumn{1}{c}{(NGC)} &
 \multicolumn{1}{c}{(kms$^{-1}$)} &
 \multicolumn{1}{c}{\null} &
 \multicolumn{1}{c}{(kms$^{-1}$kpc)} &
 (km$^2$s$^{-2}$) \\

\hline

1851 & (134.8,11.6,28.6) & 0.91 & (1493,-178,230) & -134706 \\
\hline
1904 & (46.5,-2.9,-21.5) & 0.93 & (1477,51,155) & -137390 \\
\hline
2298 & (-96.1,41.3,-57.7) & 0.79 & (949,-648,317) & -140391 \\
\hline
2808 & (-152.9,-35.5,-3.7) & 0.86 & (1038,394,35) & -152947 \\
\hline
5286 & (-202.3,42.4,-58.3) & 0.84 & (856,-366,148) & -153940 \\
\hline
6864 & (-113.0,-27.6,24.1) & 0.83 & (1144,316,324) & -143397 \\
\hline
6779 & (159.4,19.9,-76.9) & 0.86 & (677,-182,199) & -159799 \\
\hline
7089 & (231.3,24.1,28.0) & 0.88 & (1368,-192,309) & -139217 \\
\hline
\hline
362 & (147.1,7.9,-33.5) & 0.85 & (837,-57,317) & -159510 \\
\hline
1261 & (-113.8,30.5,7.2) & 0.86 & (1474,-393,351) & -132973 \\
\hline

   \end{tabular}
 \end{center}
  \label{table:params}
\end{table}

Of the 15 high energy old halo clusters, there are 6 with high
vertical action ($J_z \gtrsim 1000$ km\,s$^{-1}$ kpc). They lie well
apart from the main group. They have a wide spread in azimuthal
($J_{\phi}$) and radial ($J_R$) actions, similar to the YHs suggesting
an accretion origin. The main group are concentrated at large $J_R$,
low $J_z$ and low $J_{\phi}$ region in the action space, indicating
radial orbits.  They show surprisingly low vertical action ($J_z
\lesssim 500$ km\,s$^{-1}$) -- they are actually less extended in
$J_z$ than the low energy old halo clusters and much less extended
than the MSTO stars with similar energy.  This tight concentration,
especially in $J_z$, is interesting since the range of $J_z$ becomes
wider as we move to higher energy, as is demonstrated by the MSTO
sample.  The 8 high energy OHs forming this main group (NGCs 1851,
1904, 2298, 2808, 5286, 6864, 6779 and 7089) are marked with black
circles in Figs.~\ref{fig:actionspace} and \ref{fig:acttwo}.  For this
group of 8, the maximum $J_z$ is $\sim 360$ km\,s$^{-1}$ kpc, the
maximum $|J_\phi|$ is $\sim 500$ km\,s$^{-1}$ kpc, while the {\it
  minimum} $J_R$ is $\sim 700$ km\,s$^{-1}$ kpc. In action space,
their distribution is highly flattened and
sausage-like. Interestingly, there are no old halo clusters with
comparable energy that have high vertical action $J_z$ (see e.g., the
middle panel of Fig.~\ref{fig:acttwo}).  \citet{MG04} suggest that
15--17 per cent of the old halo clusters might have been accreted.  In
our picture, at least 8 Sausage GCs (or 14 including those with very
high $J_z$) out of 53 are accreted, in rough accord with the estimate.

The young halo globular clusters all have $E > \Ecrit$, and show a
broad spread in all actions. They include extreme prograde and
retrograde members in the sample, as well as the ones with largest
radial $J_R$ and vertical $J_z$ actions (excluding the Sgr GCs).  The
2 black open squares in Figs.~\ref{fig:actionspace} and
\ref{fig:acttwo} provide an extended selection to the Sausage
GCs. They are 2 young halo globular clusters (NGC 362, and NGC 1261)
with a rather similar horizontal branch morphology to old halo
clusters (see later) that also have similar actions and energy to the
Sausage GCs.  These are possibles rather than probables.

The upper panel of Fig.~\ref{fig:ecc} shows the apocentres and
pericentres of the sample, with lines of constant eccentricity
superposed. The \citet{GaDR2} already noted the tendency for GCs with
larger apocentres to have larger eccentricities. The 8 probable and 2
possible Sausage GCs are denoted by black open circles and open
squares.  They form a clump concentrated at high $J_R$, low $J_z$ and
low $J_{\phi}$ and they all have high orbital eccentricity $\gtrsim
0.80$.  We can also see that most of the bulge/disc clusters have low
eccentricity.  There are also many old halo clusters with comparably
low eccentricity. The young halo clusters are again widely dispersed,
as they have high energy and highly spread actions.

Finally, we must consider whether selection effects could cause
this. As the \citet{GaDR2} point out, GCs with high energy are more
likely to be observed if they are on eccentric orbits.
Even so, the middle panel of Fig.~\ref{fig:acttwo} demonstrates that
there are no old halo clusters in this energy range that have high
$J_z$. By taking the positions of GCs in our sample, and sampling
their velocities from a Gaussian with $\sigma = 130$ kms$^{-1}$, we
show the expected distribution in action space in the lower panel of
Fig.~\ref{fig:ecc}.  Notice that there is only a very mild bias
towards low $J_z$, so {\it Gaia} should have seen any high $J_z$ GCs
at this energy range, if they existed.

\section{The Sausage Globular Clusters}

The properties of the 8 probable and 2 possible Sausage GCs in energy
and action space are listed in Table~\ref{table:params}.

The identification of the Gaia Sausage in main-sequence turn-off stars
is most evident in velocity space. \cite{Be18} show that the velocity
anisotropy parameter $\beta_{\rm MSTO}$ is very extreme,
\begin{equation}
  \beta_{\rm MSTO} = 1 - {\sigma_{v_\theta}^2 + \sigma_{v_\varphi}^2
    \over 2 \sigma_{v_r}^2} \approx 0.9,
\end{equation}
Here, $v_\varphi$ is the azimuthal velocity in the direction of the
Milky Way's rotation, $v_\theta$ is increasing towards the Milky Way's
north pole and $v_r$ is the radial velocity in spherical coordinates.
Given that the $\beta = 1$ implies that all orbits are linear straight
lines through the Galactic Centre, then the metal-rich local halo
stars are very radially anisotropic. This gives the Sausage its name,
as the structure looks highly non-Gaussian and sausage-shaped in
velocity space. Fig.~\ref{fig:velocities} shows the velocities of the
GCs resolved with respect to spherical polar coordinates. The Sausage
GCs have an even more extreme value of the anisotropy parameter than
the Sausage MSTOs, with $\beta_{\rm GCs} \approx 0.95$.

\begin{figure}
  \includegraphics[width=70mm]{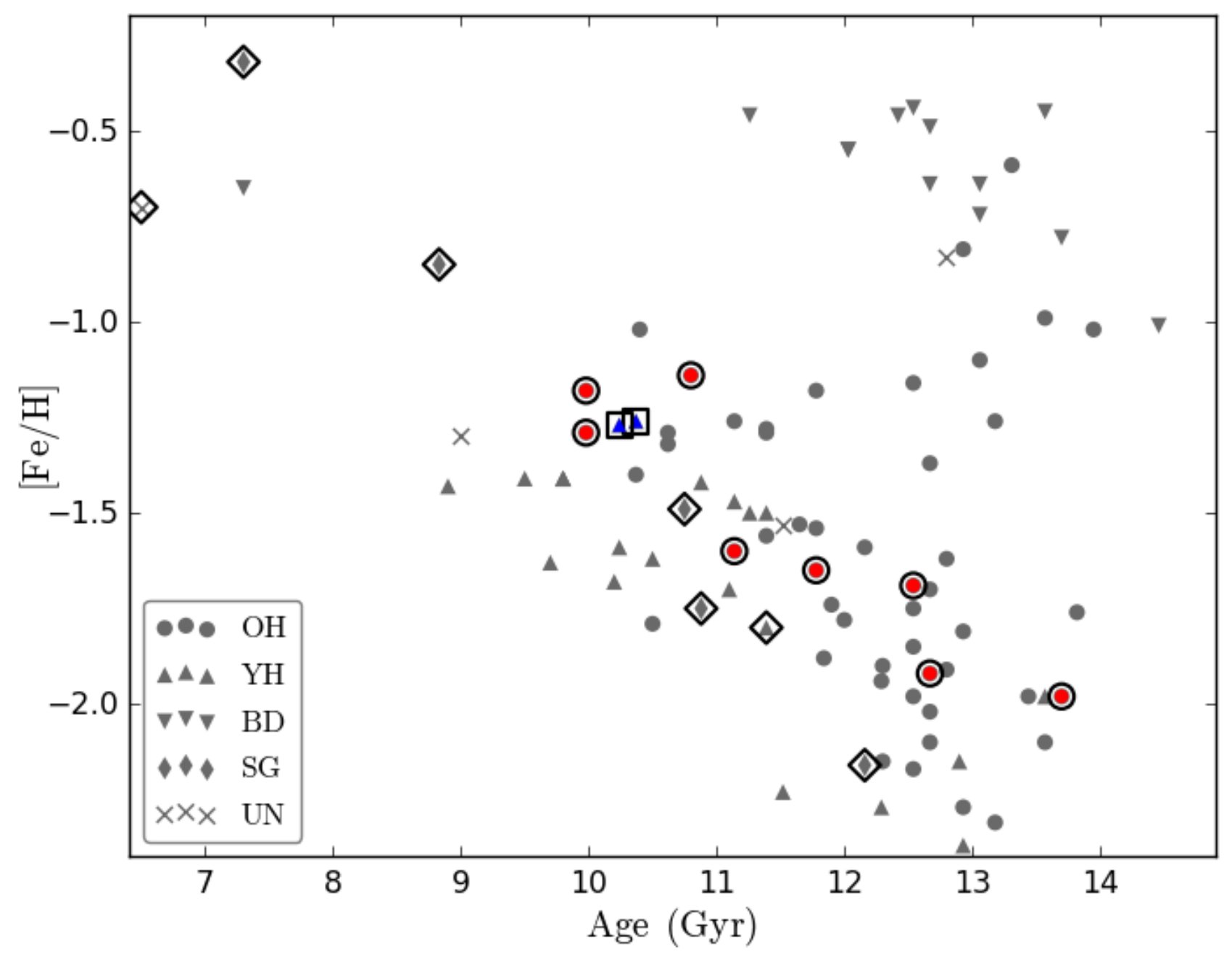}
  \includegraphics[width=70mm]{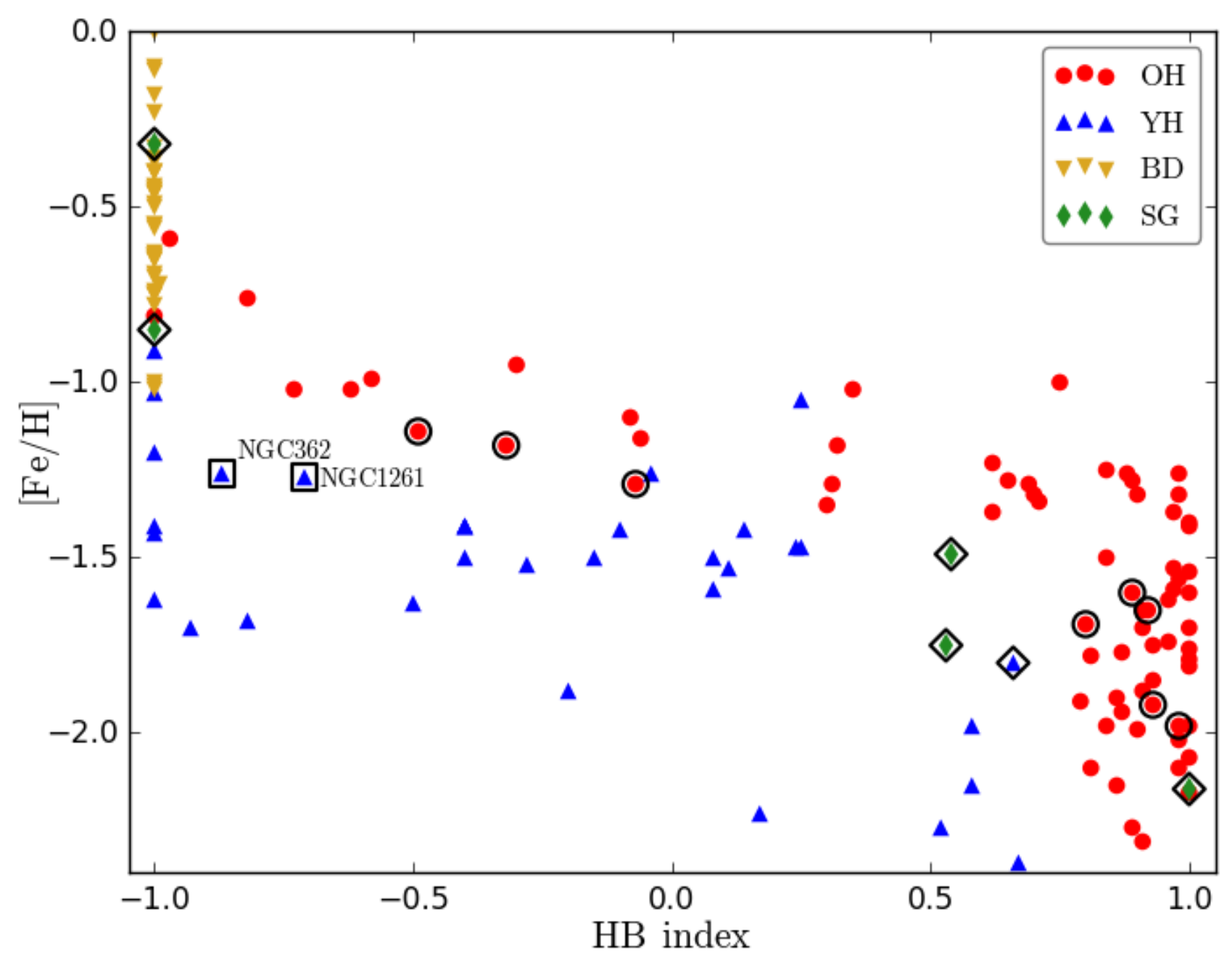}
  \caption{Upper panel: Plot of the age of GCs versus metallicity
    using data from \citet{Fo10}. The Sausage GCs are shown with
    circular (probables) and square (possibles) black boundaries. 7
    GCs that are claimed former denizens of the Sagittarius (Sgr)
    dwarf are shown as unfilled black diamonds. The sequences of Sgr
    GCs and Sausage GCs lie on two distinct, though closely matched,
    tracks. They are different from the bulk of the Milky Way GCs
    which show a constant age of $\sim 13$ Gyr independent of
    metallicity. Lower panel: Plot of horizontal branch morphology
    versus metallicity using data from \citet{MB05}. The locations of
    the two young halo clusters (NGC 362, NGC1261) are
    close to the boundary and their designation is open to debate. We have
    included them in our extended sample of Sausage GCs, as they are
    kinematically similar.
\label{fig:smallfigs}}
\end{figure}

The upper panel of Fig.~\ref{fig:smallfigs} shows age versus
metallicity for the Sausage GCs, as well as 7 GCs that have been
claimed as associates of the the Sagittarius (Sgr), specifically
Terzan 7, Terzan 8, Arp 2, Pal 12, NGC 4147, NGC 6715 and Whiting 1
\citep{Fo10}.  As noted by \citet{Fo10}, the age-metallicity relation
for the Milky Way's GCs reveals two distinct tracks. There is broad
swathe of bulge/disc and old halo globular clusters with a roughly
constant old age of $\sim 12.8$ Gyr. This comprises the bulk of the
sample. However, \citet{Fo10} pointed out that the Sgr GCs form a
separate track that branches to younger ages, and is shown as open
diamonds in Fig.~\ref{fig:smallfigs}. We find that the Sausage GCs
similarly follow a track that is very different from the bulk of the
Milky Way's {\it in situ} GCs. It is similar to, but vertically offset
from, the Sgr track.  The lower panel of Fig.~\ref{fig:smallfigs}
shows the horizontal branch index versus metallicity using data from
\citet{MB05}. The plot emphasises the ambiguous nature of the two
clusters, NGC 362, NGC 1261. Although \citet{MB05}
classified them as young halo clusters based on their horizontal
branch morphology, they are in fact close to the dividing line. We
therefore suggest that this classification can be debatable. They
are kinematically close to the Sausage GCs, who may well be their true
brethren.

\section{Discussion}

This {\it Letter} argues that there are at least eight and possibly
ten halo globular clusters that belong to a single, ancient massive
merger event identified by \citet{Be18} and responsible for the Gaia
Sausage in velocity space.

The evidence is threefold. First, there is a strong prior expectation
of finding a population of radially anisotropic GCs.  Evidence for a
major accretion event is provided by studies of the kinematics of halo
main-sequence turn-off stars in the SDSS-{\it Gaia}
catalogue~\citep{Be18,My18a}, as well as in {\it Gaia}
DR2~\citep{Ha18}. It explains the peculiar, highly non-Gaussian,
radial anisotropic local velocity distribution of halo stars (hence
the ``Gaia Sausage'').  The existence of the Sausage GCs supports the
idea of a single event and allows us to put estimates on the mass of
the progenitor. Judging from GC numbers, it must have been more
massive than Fornax and comparable to the Sgr progenitor, which
\citet{Gi17} estimated as $5 \times 10^{10} \Msun$ in total mass.
This is in good agreement with the mass estimate from simulations
already provided in \citet{Be18}.

Secondly, just as the GCs associated with the Sgr can be identified by
their agglomeration in action space, so can the GCs associated with
the ``Gaia Sausage''. A critical energy separates the young halo
clusters (which have all been accreted) from the bulge/disc clusters
(which are all formed {\it in situ}).  The old halo clusters are
mainly formed {\it in situ}, though \citet{MG04} suggest that 15--17
per cent were accreted. They straddle the critical energy.  Eight of
the old halo clusters with $E>\Ecrit$ form a narrow, clumped and
compact distribution in action space. They have characteristic low
vertical ($J_z$) and high radial ($J_R$) action. They show strong
radial anisotropy ($\beta \approx 0.95$) and highly radial, eccentric
orbits ($e \gtrsim 0.80$). These are exactly the characteristics expected
for the Sausage GCs. There may even be 2 further members -- if we, for
example, permit the inclusion of young halo clusters.

Thirdly, the 8 globular clusters identified as belonging to the ``Gaia
Sausage'' were chosen without any regard to their age or
metallicity. However, these 8 clusters show the typical
age-metallicity trend expected from dwarf spheroidals, which is
additional evidence supporting their extragalactic origin.  The time
of infall can also be roughly reckoned from the tracks in
age-metallicity space as $\sim 10$ Gyrs or $z \sim 3$,
in accord with the estimate already provided in \citet{Be18}.

Could this peculiarity of the data be due to a selection effect,
against which the \citet{GaDR2} already caution? High energy GCs are
more likely to be observed if they are on eccentric orbits.  In action
coordinates, the eccentricity roughly scales like $(J_R +
J_z)/|J_\phi|)$. So, an eccentric orbit could have large $J_R$, but it
could also have large $J_z$. However, there are no old halo clusters
in this energy range that have high $J_z$ (as the middle panel of
Fig.~\ref{fig:acttwo} demonstrates). Nonetheless, if such GCs existed,
{\it Gaia} should have identified them (see the lower panel of
Fig.~\ref{fig:ecc}).  This argues against the emptiness of the high
$J_z$ portion of action space being a mere selection effect.


\end{document}